\documentclass{article}
\usepackage{a4wide}
\usepackage{natbib} 					
\usepackage{graphicx}
\usepackage{amsfonts}
\usepackage{amsmath}
\usepackage{lscape}
\usepackage{booktabs}
\usepackage[colorlinks]{hyperref} 	
\hypersetup{  
citecolor=blue, 						
linkcolor= blue} 						

\begin{document}
\title{Fractal Markets Hypothesis and the Global Financial Crisis: Scaling, Investment Horizons and Liquidity}
\date{}
 \author{
 Ladislav Kristoufek \thanks{Institute of Economic Studies, Faculty of Social Sciences, Charles University, Opletalova 26, 110 00, Prague, Czech Republic,
Institute of Information Theory and Automation, Academy of Sciences of the Czech Republic, Pod Vodarenskou Vezi 4, Prague 8, 182 08, Czech Republic}
 }
 
\maketitle

\begin{abstract}

We investigate whether fractal markets hypothesis and its focus on liquidity and investment horizons give reasonable predictions about dynamics of the financial markets during the turbulences such as the Global Financial Crisis of late 2000s. Compared to the mainstream efficient markets hypothesis, fractal markets hypothesis considers financial markets as complex systems consisting of many heterogenous agents, which are distinguishable mainly with respect to their investment horizon. In the paper, several novel measures of trading activity at different investment horizons are introduced through scaling of variance of the underlying processes. On the three most liquid US indices -- DJI, NASDAQ and S\&P500 -- we show that predictions of fractal markets hypothesis actually fit the observed behavior quite well.

\end{abstract}



\normalsize
\section{Introduction}

Efficient markets hypothesis (EMH) has been a hot topic since its introduction in 1960s \citep{Fama1965,Fama1970,Samuelson1965}. For its simplicity and intuitive logical structure, EMH has been widely accepted as a cornerstone of the modern financial economics. Since the very beginning, EMH has been criticized on several fronts, mainly theoretical -- that it is only a set of practically meaningless tautologies \citep{LeRoy1976} -- and empirical -- that it is frequently violated and financial markets are at least partially predictable \citep{Malkiel2003}.

In his pioneering paper, Fama \citep{Fama1970} describes the efficient market as the one where all available information are already reflected in the asset prices. In his later work \citep{Fama1991}, he defined the efficient market through the language of mainstream economics as the one where prices reflect available information to the point where marginal gain from using the information equals marginal cost of obtaining it. Based on this assertion, the efficient market is defined as a random walk \citep{Fama1965}, contrary to Samuelson's formulation through a martingale \citep{Samuelson1965}. Either way, the efficient market leads to a Brownian motion of the asset prices, i.e. a process with independent and identically normally distributed increments. Apart from uncorrelatedness of the increments (autocorrelations have been shown to vanish for lags higher than units of minutes \citep{Stanley1999}), the implications of EMH have been widely rejected in empirical studies \citep{Cont2001}. However, the most severe shortcoming of EMH is its ignorance to extreme (and sometimes devastating) events on the capital markets, which theoretically should have never happened \citep{Stanley2003}.

EMH has far-reaching implications, which are discussed in majority of financial economics textbooks \citep{Elton2003} -- investors are rational and homogeneous, financial returns are normally distributed, standard deviation is a meaningful measure of risk, there is a tradeoff between risk and return, and future returns are unpredictable. To some extent, all of these implications can be easily attacked with empirical analysis. For our purposes, the first implication of homogeneous investors is crucial. It implies that all the investors use the available information in the same way and thus they operate on the same investment horizon (or theoretically the same set of investment horizons). However, it is known that capital markets comprise of various investors with very different investment horizons -- from algorithmically-based market makers with the investment horizon of fractions of a second, through noise traders with the horizon of several minutes, technical traders with the horizons of days and weeks, and fundamental analysts with the monthly horizons to pension funds with the horizons of several years. For each of these groups, the information has different value and is treated variously. Moreover, each group has its own trading rules and strategies, while for one group the information can mean severe losses, for the other, it can be taken a profitable opportunity. This environment creates a very complex system, which can be hardly described by oversimplified EMH.

On contrary, fractal markets hypothesis (FMH) \citep{Peters1994} has been constructed based on the most general characteristics of the markets. In its core, it is based on a notion completely omitted in EMH -- liquidity. According to FMH, liquidity provides smooth pricing process in the market, making it stable. If liquidity ceases, market becomes unstable and extreme movements occur. In the literature, FMH is usually connected with detection of fractality or multifractality of the price processes of financial assets \citep{Peters1994,Onali2009,Onali2011}. However, it has not been put to test with respect to its predictions about causes and implications of critical events in the financial markets. In this paper, we analyze whether these predictions fit the observed behavior in the stock markets before and during the current Global Financial Crisis (2007/2008--?). Mainly, we are interested in the behavior of investors at various investment horizons as well as in scaling of the market returns. To do so, we utilize a sliding window estimation of generalized Hurst exponent $H(q)$ with $q=2$ (usually called local or time-dependent Hurst exponent). Moreover, we introduce several new measures of trading activity at different investment horizons based on decomposition of Hurst exponent and variance scaling.

The local Hurst exponent approach has been repeatedly used to analyze potential turning and critical points in the stock market behavior. \cite{Grech2004} studied the crashes of 1929 and 1987 focusing on behavior of Dow Jones Industrial Index and showed that the local Hurst exponent analysis can provide important signals about coming extreme events. In the series of papers, \cite{Grech2008,Czarnecki2008} studied the critical events of the Polish main stock index WIG20 and again presented the local Hurst exponent as a useful tool for detection of coming crashes (together with log-periodic model of \cite{Sornette1996}). \cite{Domino2011,Domino2012} further studied the connection between local Hurst exponent behavior and critical events of WIG20 index. \cite{Kristoufek2010} applied the similar technique on detection of coming critical points of PX50 index of the Czech Republic stock market and uncovered that the functioning is very similar. \cite{Morales2012} broadened the application of time-dependent Hurst exponent on a wide portfolio of the US stocks and showed that the values of Hurst exponent can be connected to different phases of the market. 

In this paper, we show that behavior of the time-dependent Hurst exponent is connected to various phases of the market. Moreover, we uncover that there are some common patterns before the critical points. Most importantly, the Global Financial Crisis is detected to be connected with unstable trading and unbalanced activity at different investment horizons which is asserted by FMH. The paper is organized as follows. In Section 2, we give basic definitions of fractal markets hypothesis. Section 3 describes multifractal detrended fluctuation analysis, which we use for the generalized Hurst exponent estimation, and introduces new measures of trading activity at specific investment horizons. In Section 4, we test whether the assertions of FMH are actually observed in the real market. All three analyzed indices -- DJI, NASDAQ and S\&P500 -- share several interesting patterns before and during the current financial crisis, which are in hand with FMH.

\section{Fractal markets hypothesis}

Fractal markets hypothesis (FMH) was proposed by \cite{Peters1994} as a follow-up to his earlier criticism of EMH \citep{Peters1991}. The cornerstone of FMH is a focus on heterogeneity of investors mainly with respect to their investment horizons. The market consists of the investors with investment horizon from several seconds and minutes (market makers, noise-traders) up to several years (pension funds). Investors with different investment horizons treat the inflowing information differently and their reaction is correspondingly distinct (market participants with short investment horizon focus on technical information and crowd behavior of other market participants, whereas investors with long investment horizon base their decisions on fundamental information and care little about crowd behavior). Specific information can be a selling signal for a short-term investor but an opportunity to buy for a long-term investor and vice versa. The existence of investors with different horizons assures a stable functioning of the market. When one horizon (or a group of horizons) becomes dominant, selling or buying signals of investors at this horizon will not be met with a reverse order of the remaining horizons and prices might collapse. Therefore, the existence and activity of investors with a wide range of investment horizons is essential for a smooth and stable functioning of the market \citep{Rachev1999,Weron2000}.

Fractal markets hypothesis thus suggests that during stable phases of the market, all investment horizons are equally represented so that supply and demand on the market are smoothly cleared. Reversely, unstable periods such as "crises" occur when the investment horizons are dominated by only several of them so that supply and demand of different groups of investors are not efficiently cleared. This two implications give us characteristic features to look for in the market behavior.

FMH is tightly connected to a notion of multifractality and long-range dependence in the underlying series. Process $X_t$ is considered multifractal if it has stationary increments which scale as

\begin{equation*}
\langle|X_{t+\tau}-X_t|^q\rangle\propto \tau^{qH(q)}
\label{eq1}
\end{equation*}

for integer $\tau>0$ and for all $q$ \citep{Calvet2008}. $H(q)$ is called generalized Hurst exponent and its dependence on $q$ separates the processes into two categories -- monofractal (or unifractal) for constant $H(q)$ and multifractal when $H(q)$ is a function of $q$. For $q=2$, we consider long-range dependence of the increments of the process $X_t$. As this case is the most important for us as it characterizes scaling of variance (and we treat variance as a sign of a trading activity), we label $H \equiv H(2)$ further in the text. Hurst exponent $H$ is connected to asymptotically hyperbolically decaying autocorrelation function $\rho(k)$, i.e. $\rho(k) \propto k^{2H-2}$ for $k \rightarrow \infty$. For $H=0.5$, we have a serially uncorrelated process; for $H>0.5$, we have a persistent process; and for $H<0.5$, we an anti-persistent process. Persistent processes are visually trending yet still remain stationary, whereas anti-persistent processes switch their sign more frequently then random processes do.

\section{Scaling of stock returns}

In this section, we present the method we use for the estimation of generalized Hurst exponent -- multifractal detrended fluctuation analysis (MF-DFA) -- and several novel measures connected to a trading activity at various trading horizons. MF-DFA is applied here because it is standardly used in the local Hurst exponent literature \citep{Grech2004,Grech2008,Czarnecki2008,Kristoufek2010} and compared to other methods, such as generalized Hurst exponent approach \citep{DiMatteo2005,DiMatteo2007,Barunik2010,Kristoufek2011}, it provides wider range of scales to analyze. As we want to compare as many investment horizons as possible, such a distinction leads to MF-DFA.

\subsection{Multifractal detrended fluctuation analysis}

Multifractal detrended fluctuation analysis (MF-DFA) is a generalization of detrended fluctuation analysis (DFA) of \cite{Peng1993,Peng1994}. \cite{Kantelhardt2002} proposed MF-DFA to analyze scaling of all possible moments $q$, not only the second one ($q=2$) as for DFA. One of the advantages of MF-DFA and DFA over other techniques is that it can be applied on series with $H>1$, i.e. a higher order of integration.

In the procedure, one splits the series of length $T$ into segments of length $s$. For each segment, a polynomial fit $\widehat{X_{s,l}}$ of order $l$ is constructed for the original segment $X_{s}$. In our analysis, we apply a linear fit so that $l=1$ and we will omit the label onwards. Note that the filtering procedure can be chosen not only from polynomial fits but also from moving average, Fourier transforms and various others \cite{Kantelhardt2009}. Detrended signal is constructed for each segment as $Y_s=X_s-\widehat{X_s}$. Fluctuation $F^2_{DFA,q}(i,s)$ is defined for each sub-period $i$ of length $s$ as $$F^2_{DFA,q}(i,s)=\left(\sum_{i=1}^{[T/s]}{Y_{i,s}^2}/s\right)^{\frac{1}{2}}.$$ As $T/s$ is not necessarily an integer, we calculate the fluctuations in the segments starting from the beginning as well as from the end of the series not to omit any observation. By doing so, we obtain $2[T/s]$ fluctuations $F^2_{DFA,q}(i,s)$. The fluctuations are then averaged over all segments with length $s$ to obtain the average fluctuation $$F_{DFA,q}(s)=\left(\sum_{j=1}^{2[T/s]}{F^2_{DFA,q}(j,s)}/2[T/s]\right)^{\frac{1}{q}}.$$ The average fluctuations scale as $F_{DFA,q}(s)= cs^{H(q)}$ where $H(q)$ is a generalized Hurst exponent and $c$ is a constant. For $q=2$, we obtain standard DFA for a long-range dependence analysis. Hurst exponent is usually estimated only for a range of scales $s$ between $s_{min}$ and $s_{max}$. The minimum scale is set so that the fit in each segment can be efficiently calculated and the maximum scale is set so that the average fluctuation for this scale is based on enough observations.

\subsection{Scaling-based liquidity measures}

Estimation of Hurst exponent compresses all the information from the dynamics of the process into a single value. However, the procedure can be also decompressed to give us some additional information. From the economic point of view, the segment's length $s$ can be taken as a length of an investment horizon. Fluctuation corresponding to the horizon $s$ can be then taken as a proxy for activity of traders with a horizon of $s$. From our previous discussion about situations of market instabilities in FMH framework, we propose several new measures.

Trading activity of investors with very short investment horizons can be approximated with $\widehat{F(0)}=e^{\widehat{c}}$, which is an estimate of fluctuation at horizon $s \rightarrow 0$. In an unstable market, it is assumed that investors at the very short horizons will be the most active ones. Also, some long-term investors might shorten their horizons \cite{Peters1994}. Therefore, we assume that close to and during market turmoils, $\widehat{F(0)}$ will increase compared to the stable periods.

In a stable market, all investment horizons are represented uniformly (or at least approximately uniformly). During critical points, the long-term investors either restrict or even stop their trading activities and the short-term investors become dominant. Trading activity and thus fluctuations $F^2$ at shorter trading horizons will be higher than rescaled trading activity at longer horizons. Therefore, Hurst exponent $H$ would be decreasing shortly before and during the turbulent times at the market. This is in hand with the definition of irregular market of \cite{Corazza2002}.

In a regular market, the scaling of variance should be stable, i.e. fluctuations $F^2$ for different horizons $s$ should lay on a straight line. If any of the investment horizons becomes dominant, the scaling would be less precise. To measure such dispersion of trading activity at different investment horizons, we introduce $F_{\sigma}$, which is a standard deviation of rescaled fluctuations, and $F_{R}$, which is a range of rescaled fluctuations: 
$$F_{\sigma}=\sqrt{\frac{\sum_{s=s_{min}}^{s_{max}}\left(F^2(s)/s^{2\widehat{H}}-\sum_{s=s_{min}}^{s_{max}}{F^2(s)/s^{2\widehat{H}}(s_{max}-s_{min}+1)}\right)^2}{s_{max}-s_{min}}};$$
$$F_{R}=\max_{s_{min}\le s \le s_{max}}\left(F^2(s)/s^{2\widehat{H}}\right)-\min_{s_{min}\le s \le s_{max}}\left(F^2(s)/s^{2\widehat{H}}\right)$$
During turbulent times, both $F_{\sigma}$ and $F_R$ are expected to increase. In a similar manner, we also define a ratio $F_r$ between rescaled fluctuations of the horizons with the maximal and minimal rescaled fluctuation. In an ideal market with uniformly represented investment horizons, we would have $F_r=1$. The further $F_r$ is from 1, the less stable is the scaling and thus also the less stable the market is.

\section{Application to the Global Financial Crisis}

\subsection{Data and methodology}

To check whether the implications of FMH hold, we apply the proposed methodology to the daily series of three US indices -- Dow Jones Industrial Average Index (DJI), NASDAQ Composite Index (NASDAQ) and S\&P500 Index (SPX) -- between the beginning of 2000 and the end of 2011. As it is widely believed that the crisis started in the USA and spilled over to the other parts of the world, we choose the US indices because they should signify the coming and continuing crisis the best. If the predictions of FMH hold, we expect local Hurst exponent to be decreasing before the critical point and remaining below $H=0.5$ during the crisis. In a similar way, 
trading activity at the short horizons $\widehat{F(0)}$ should be increasing before the crisis and remain high during the crisis compared to the more stable periods. The very same expectations hold for $F_{\sigma}$ and $F_{R}$. For $F_r$, we expect the values to be further from one before and during the crisis times.

We use a moving (sliding) window procedure to the dataset. The window length is set to $T=500$ trading days (approximately two trading years) and a step to one day. For MF-DFA, we set $s_{min}=10$ and $s_{max}=T/10=50$. This way, we can estimate $H$, $\widehat{F(0)}$, $F_{\sigma}$, $F_R$ and $F_r$ and comment on their evolution in time and during various phases of the market behavior. 

To meet stationarity condition, which is essential for correct Hurst exponent estimation, we filter the raw series with GARCH(1,1). Therefore, the analysis is made on filtered series defined as $fr_t=r_t/{\sqrt{h_t}}$, where $fr_t$ is a filtered return at time $t$, $r_t$ is a raw return at time $t$, defined as $r_t=\log(S_t/S_{t-1})$ with $S_t$ being a stock index closing value at time $t$, and $h_t$ is a conditional variance obtained from GARCH(1,1) at time $t$. The GARCH-filtering is a crucial addition to the methodology because without comparable volatility in different time windows, we would not be able to say whether e.g. an increase in $\widehat{F(0)}$ is caused by changing structure of investors activity or just an increase of variance across all scales (investment horizons).

\subsection{Results}

\begin{figure}[ht]
\center
\begin{tabular}{cc}
\includegraphics[width=2.5in]{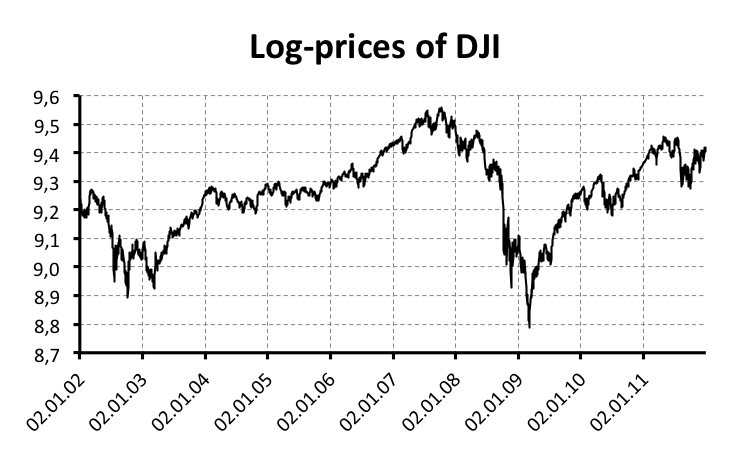}&\includegraphics[width=2.5in]{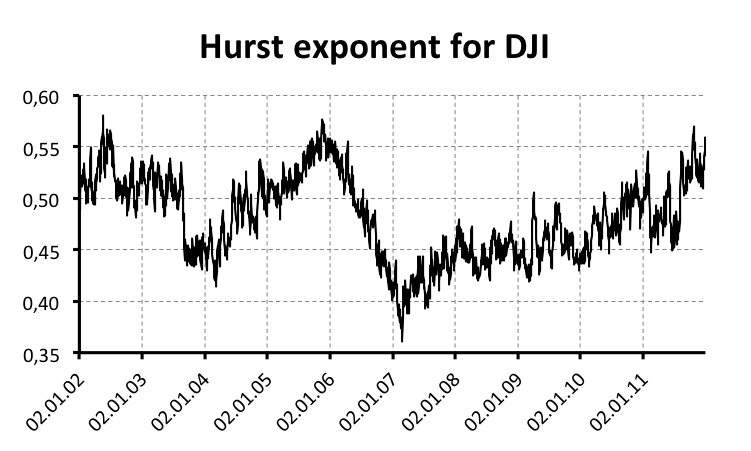}\\
\includegraphics[width=2.5in]{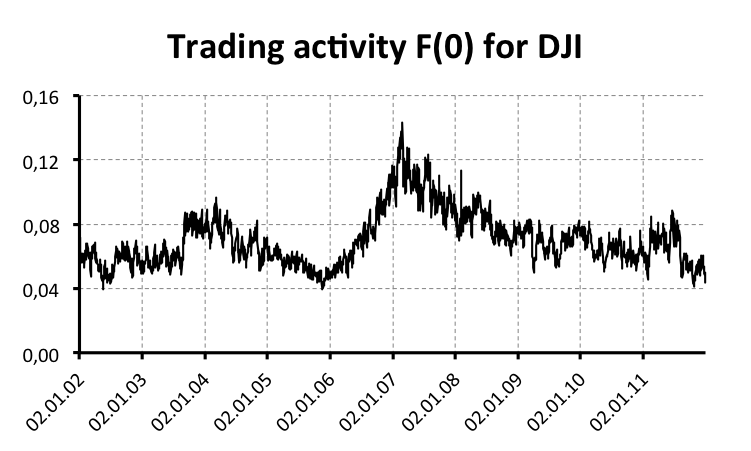}&\includegraphics[width=2.5in]{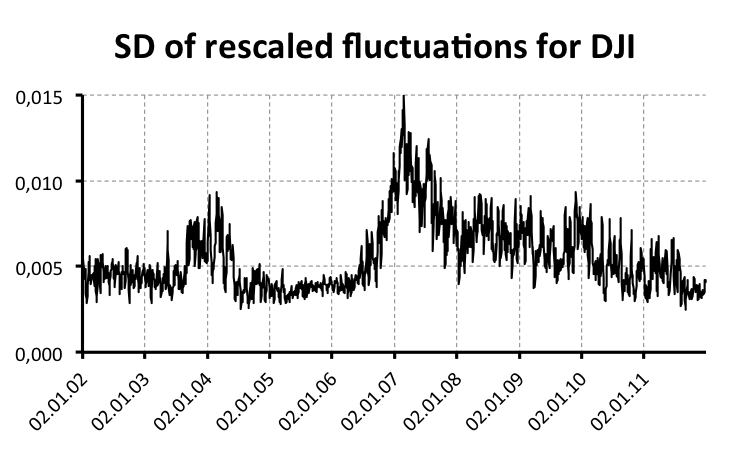}\\
\includegraphics[width=2.5in]{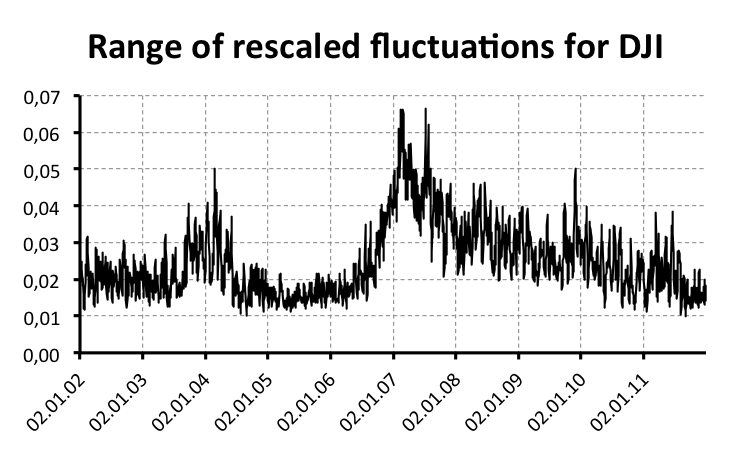}&\includegraphics[width=2.5in]{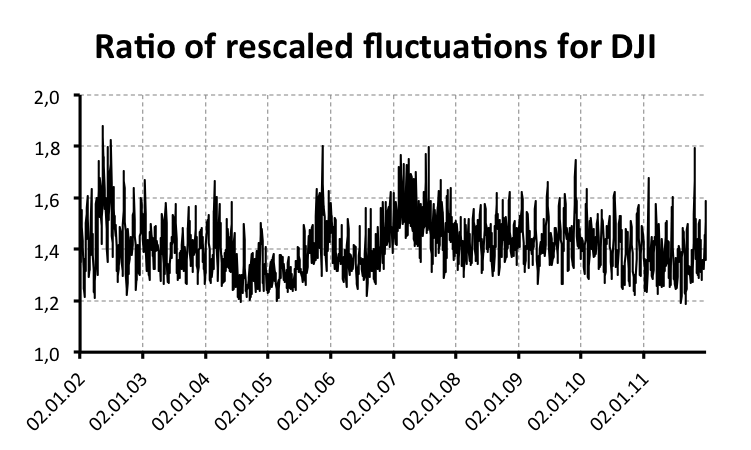}\\
\end{tabular}
\caption{Results for DJI index.\label{fig1}}
\end{figure}

\begin{figure}[ht]
\center
\begin{tabular}{cc}
\includegraphics[width=2.5in]{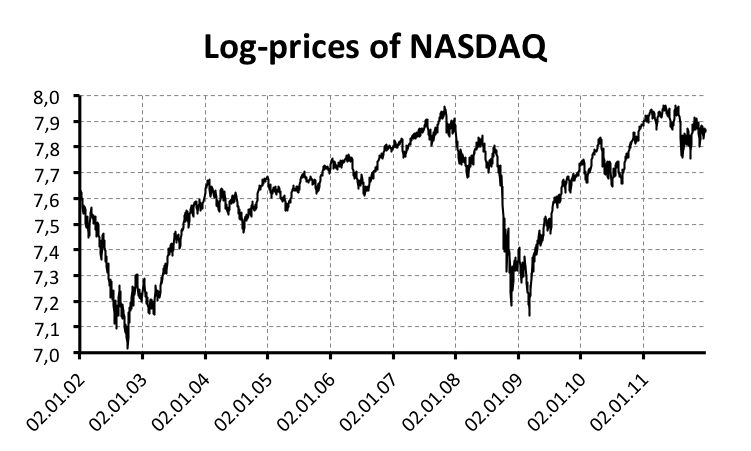}&\includegraphics[width=2.5in]{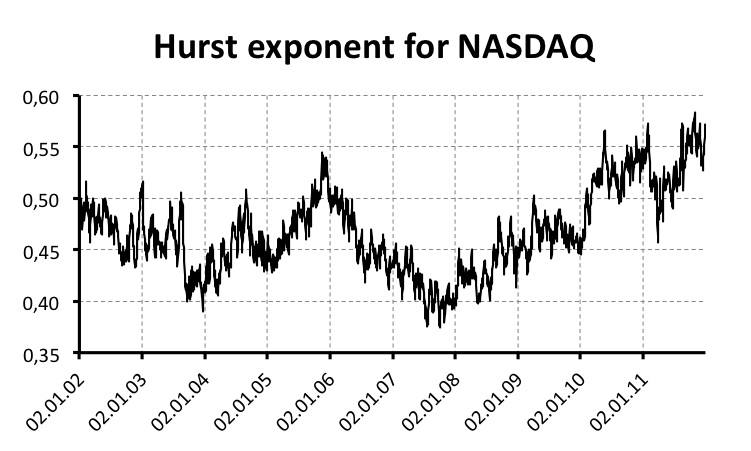}\\
\includegraphics[width=2.5in]{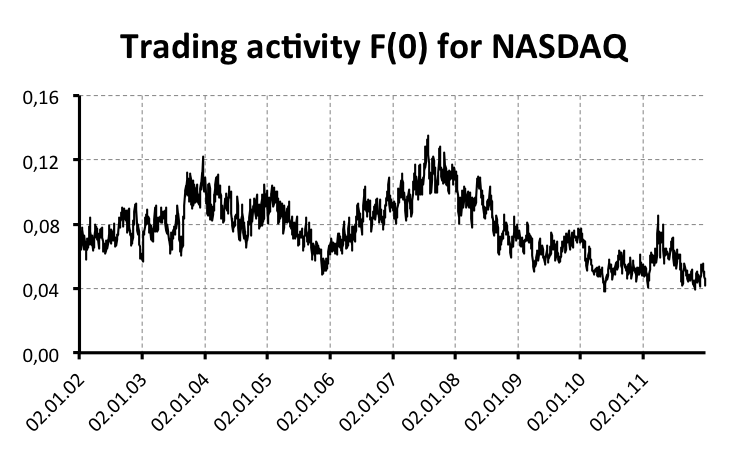}&\includegraphics[width=2.5in]{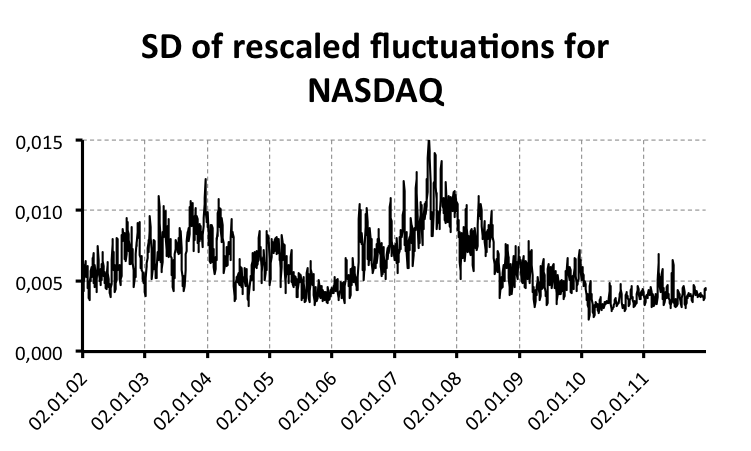}\\
\includegraphics[width=2.5in]{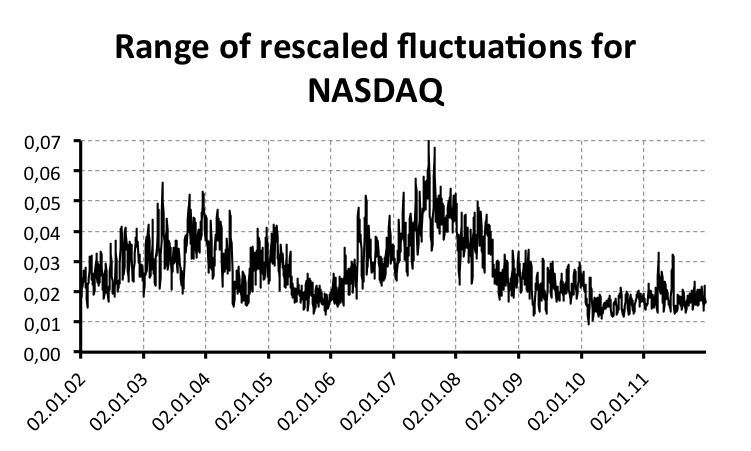}&\includegraphics[width=2.5in]{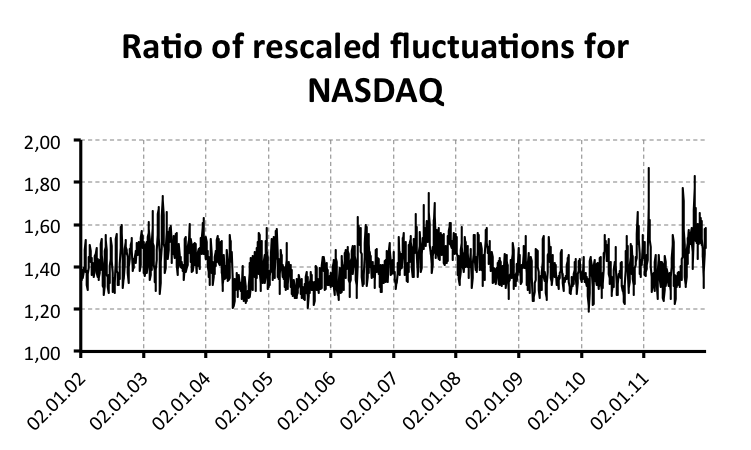}\\
\end{tabular}
\caption{Results for NASDAQ index.\label{fig2}}
\end{figure}

\begin{figure}[ht]
\center
\begin{tabular}{cc}
\includegraphics[width=2.5in]{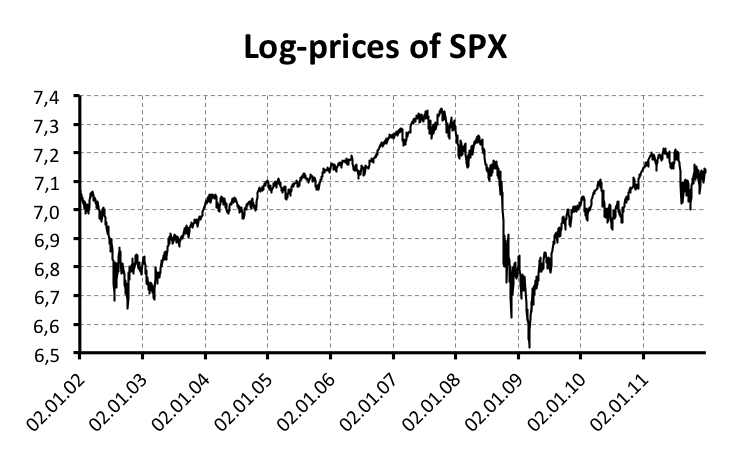}&\includegraphics[width=2.5in]{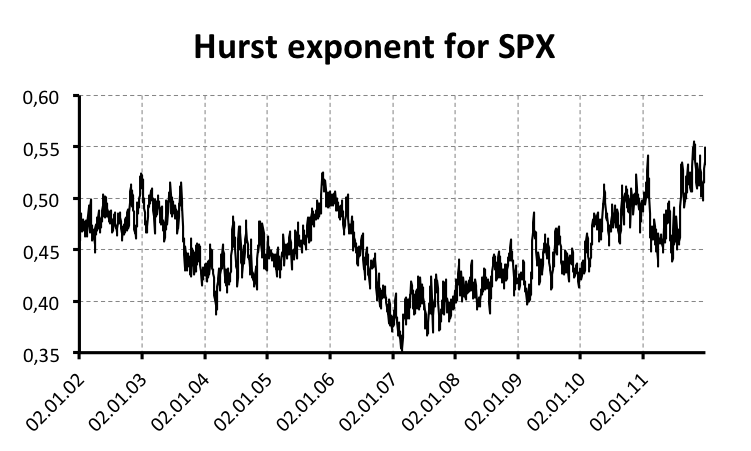}\\
\includegraphics[width=2.5in]{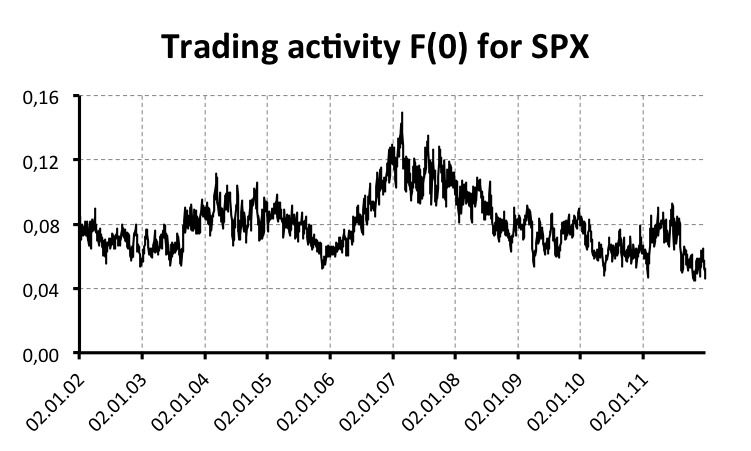}&\includegraphics[width=2.5in]{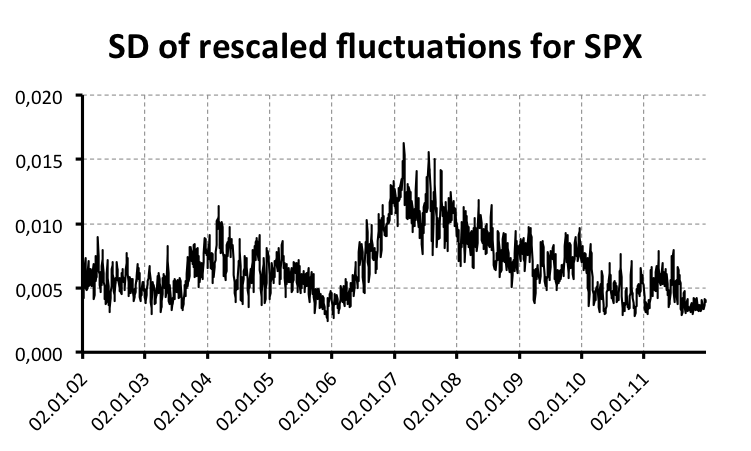}\\
\includegraphics[width=2.5in]{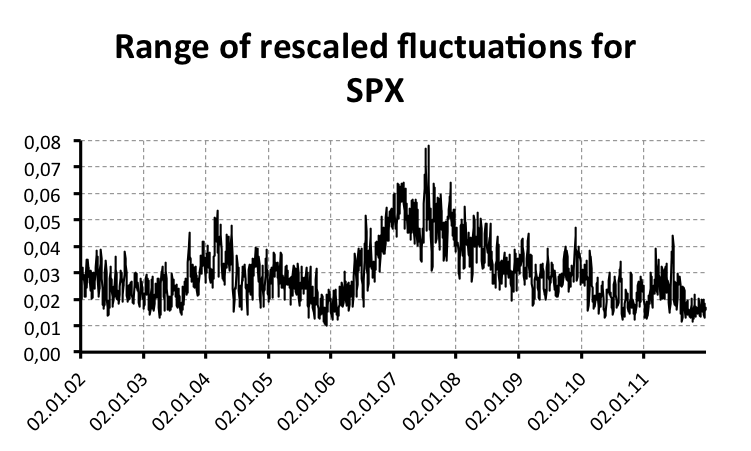}&\includegraphics[width=2.5in]{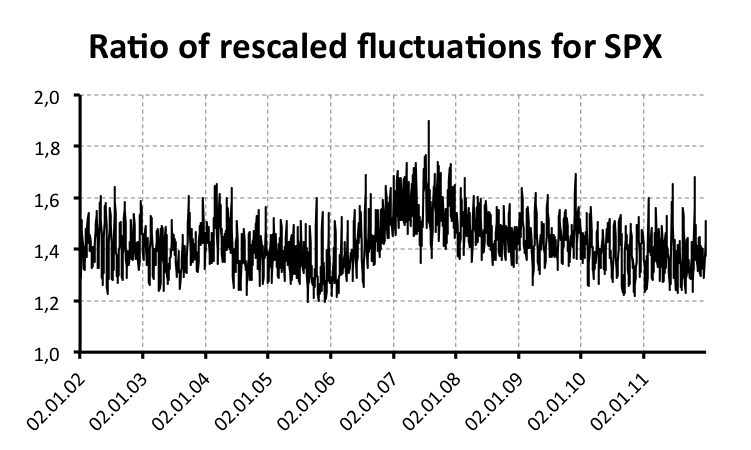}\\
\end{tabular}
\caption{Results for S\&P500 index.\label{fig3}}
\end{figure}

Results for all three analyzed indices are summarized in Figs. \ref{fig1}--\ref{fig3}. All the indices reached their post-DotCom bubble maxima in the latter half of 2007, which were followed by progressively decreasing trend culminating at the turn of 2008. Similarly to the indices all over the world, the US indices lost around half of their value during that approximately 1.5 year -- the loss accounted for 53.78\% , 61.22\%, and 56.77\% for DJI, NASDAQ and S\&P500, respectively. However, all three indices have been strongly increasing since the bottoms in the beginning of 2009 and have almost recovered all the losses by the end of 2011.

For all three indices, we observe very similar patterns in the behavior of all variables of interest. Starting with the behavior of the local Hurst exponent, we can see that for all indices, $H$ followed strong decreasing trend from the break of 2005 and 2006 till the beginning of 2007. For NASDAQ, the trend followed even to the second half of 2007. Such a behavior can be attributed to a changing structure of investors' activity -- increasingly more trading activity was taking place at short investment horizons. The end of these strong downward trends of $H$ are connected to the end of soaring gains of all the analyzed indices. For NASDAQ, the end of the local Hurst exponent trend can be even connected to attaining the maximal values in the of 2007. Afterwards, the local Hurst exponent follows a slow increasing trend for all three indices. However, $H$ remains below the value of 0.5, which is associated with a random behavior, for a rather long period. The lengths of these periods vary across the analyzed indices -- the longest for S\&P500, which is the index with the lowest gains after the crisis. These two phenomena might be connected because for NASDAQ, which has been the most increasing market of the three after the crisis, we observe the values of $H$ even above 0.5 in 2010 and 2011. Note that values of $H>0.5$ indicate dominance of long-term traders (a higher trading activity at long investment horizons) and thus a belief in good prospects of the market situation.

When we look at the short horizons trading activity $\widehat{F(0)}$, we observe that it was increasing in very similar period as Hurst exponent was decreasing the in the previous paragraph. The measure increased from values of approximately 0.04 up to over 0.12 for all three analyzed indices. After reaching its peak, the trading activity at the short horizons was slowly decreasing back to the original levels of the beginning of 2006. Again, we observe differences in the duration of this downward trend. For NASDAQ, the pre-crisis levels of short-term trading activity was reached around the beginning of 2009 and since then, the activity has remained relatively stable. On contrary, DJI has not reached the pre-crisis levels yet and S\&P500 got back to the pre-crisis levels during 2010. Even though the durations and magnitudes of short-term trading activity vary between the analyzed markets, we observe that the most critical points of the crisis were connected to increased trading activity of short-term investors.

The other two measures -- standard deviation of rescaled fluctuations $F_{\sigma}$ and range of rescaled fluctuations $F_R$ -- tell a very similar story. Since both are the measures of instability of variance scaling across different investment horizons, this is not surprising. The results are actually very alike to the dynamics of estimated fluctuations at very short investment horizons discussed in the previous paragraph -- very rapid increase starting in 2005/2006 turning which followed to the first half of 2007 (and again longer for NASDAQ). According to FMH, unevenly represented investment horizons imply complicated matching between supply and demand at the financial market. Therefore, the increasing instability of trading activity at different investment horizons indicates growing problems of this supply--demand matching. After the strong increases between 2006 and 2007, both measures started decreasing afterwards. However, only NASDAQ has recovered the pre-crisis stability levels.

The last measure we present -- the ratio between trading activity at the horizon with the highest and with the lowest activity -- uncovers quite similar results. Note that even in the calm periods before the last crisis, the ratio is not equal or close to one as it theoretically should be (for a perfectly scaling variance). Between the beginning of 2006 and the first half of 2007, the ratio increased from around 1.2 up to 1.6 for all three indices. Notably, the following decreasing trend was the fastest for NASDAQ while it took much longer to DJI and S\&P500 to recover. However, the results for this last measure are probably the weakest as the measure is very volatile in time.

\section{Conclusions and discussion}

Efficient market hypothesis is unable to describe the behavior of the financial markets during the last (current) crisis starting in 2007/2008 in a satisfying way. We analyze whether an alternative approach -- fractal markets hypothesis -- gives more reasonable predictions. The cornerstone of FMH is liquidity connected to the trading activity at different investment horizons. If the investors with different horizons are uniformly distributed across scales, supply and demand for financial assets work efficiently. However, when a specific investment horizon (or a group of horizons) starts to dominate the situation in the market, the supply--demand matching ceases to work and a critical point emerges. To test whether this crucial assertion of FMH holds for the current crisis, we used the local Hurst exponent approach as well as the introduced set of new measures of trading activity based on Hurst exponent decomposition.

We found that the behavior at various investment horizons is quite well described by the FMH before and during the current Global Financial Crisis. Analyzing three stock indices of the USA -- DJI, NASDAQ and S\&P500 -- we showed that the local Hurst exponent decreases rapidly before the turning of the trend, which is in hand with previously published results \citep{Grech2004,Grech2008,Czarnecki2008,Kristoufek2010}. Moreover, with a use of the new measures of trading activity, we uncovered that investors' trading activity indeed changes before and during the crisis period compared to the preceding stable periods. Before the crisis, the structure of trading activity at different investment horizons changed remarkably with rapidly increasing activity at the shortest horizons, i.e. short-term investors started to dominate and long-term investors showed no faith in a continuing growth. Also, the stability of investment horizons representation changed before the current turbulent times. Uniformity of investment horizons representation started to cease before an outburst of the crisis. During and after the most severe losses, the indicators started to stably return to the pre-crisis levels. However, they fully recovered only for NASDAQ index while DJI and S\&P500 are just attaining the former stability. Note that NASDAQ, which is the index with the fastest recovering investment horizons measures, is also the index which returned to the pre-crisis values the fastest.

Summarizing, we have showed that fractal markets hypothesis gives reasonable predictions of market dynamics in the turbulent times. Trading activity at various investment horizons ensuring efficient clearing of supply and demand in the market, which guarantees high liquidity, turns out to be a crucial attribute of a well-functioning and stable market.

\section*{Acknowledgements}

The support from the Grant Agency of Charles University (GAUK) under project $118310$, Grant Agency of the Czech Republic (GACR) under projects 402/09/0965 and P402/11/0948, and project SVV 261 501 are gratefully acknowledged.

\bibliographystyle{chicago}
\bibliography{FMH}

\end{document}